# AI Combines, Humans Socialise: A SECI-based Experience Report on Business Simulation Games

*Corresponding author:*

Nordine Benkeltoum
https://orcid.org/0000-0001-8024-3165
Lecturer in Management Information Systems
Centrale Lille Institut
CS 20048
59651 Villeneuve d'Ascq cedex
France
nordine.benATgmail.com

# Abstract

*Background.* **Business Simulation Games** (**BSG**) are widely used to foster experiential learning in complex managerial and organisational contexts by exposing students to decision-making under uncertainty. In parallel, **Artificial Intelligence** (**AI**) is increasingly integrated into higher education to support learning activities. However, despite growing interest of **AI** in education, its specific role in BSG and its implications for knowledge creation processes remain under-theorised.

*Intervention.* This paper reports on the integration of generative **AI** tools into a **BSG** designed for engineering students. **AI** was embedded as a support mechanism during the simulation to assist students in analysing events, reformulating information, and generating decision-relevant insights, while instructors retained responsibility for supervision, debriefing, and complex issues.

*Methods.* Adopting a qualitative experience-report approach, the study draws on the **SECI model** (Socialisation, Externalisation, Combination, Internalisation) as an analytical framework to examine how students and instructors interacted with **AI** during the simulation and how different forms of knowledge were mobilised and developed.

*Results.* The findings indicate that **AI** primarily supports the Combination phase of the **SECI** model by facilitating the rapid synthesis, reformulation, and contextualisation of explicit knowledge. In contrast, the processes of Socialisation, Externalisation, and Internalisation remained largely dependent on peer interaction, individual reflection, and instructor guidance.

*Discussion.* The results suggest a functional boundary in human-**AI** collaboration within simulation-based learning. **AI** acts as a cognitive enhancer that improves responsiveness and access to explicit knowledge, but it does not replace the pedagogical role of instructors in supporting the development of tacit knowledge, competencies, and phronesis.

*Conclusion.* Integrating AI into **BSG** can enhance learning efficiency and engagement, but effective experiential learning continues to rely on active human supervision. Future research should investigate instructional designs that better support tacit knowledge acquisition in **AI**-assisted simulations.

# Introduction

Artificial intelligence (AI) has widely recognised capabilities in generating structured text (Labadze et al., 2023). In several domains, such as strategy games, AI surpasses human performance (Bengio et al., 2024). AI is increasingly used in executive professional training and university curricula (Faisal et al., 2022; Humpherys et al., 2022), while a growing number of educators has an optimistic view of AI use in the classroom (Larson et al., 2024). In parallel, Business Simulation Games (BSG) have emerged as an effective pedagogical approach for enhancing student engagement, particularly in complex and dynamic learning environments (Humpherys et al., 2022). By integrating game mechanics, BSG aim to improve the learning experience while fostering active participation and sustained involvement. BSG are particularly prevalent in entrepreneurship education. Nevertheless, prior research indicates that uncertainty remains a significant limitation, as it is inherently difficult to model within simulation environments (Chen et al., 2022).

Although real-world data and case materials can be incorporated into teaching, exposing students to authentic professional situations remains challenging, particularly when addressing managerial decision-making and its consequences. Management decisions such as workforce reductions or responses to industrial accidents involve complex social, legal, and health-related implications that are difficult to simulate using traditional teaching methods. Role-play consists of exposing students to a situation in which they play a prescribed role (Hsu, 1989). Role-playing within business simulation provides an appropriate framework for creating plausible managerial scenarios and confronting students with the consequences of their decisions in a controlled environment (Faisal et al., 2022, p. 19).

During gameplay, instructor responsiveness is crucial, particularly in providing timely feedback to students (Chen et al., 2022, p. 2; Larson et al., 2024). Consequences must be communicated promptly in order to maintain immersion and reinforce learning outcomes. However, generating credible and context-specific content in near real time remains difficult without technological support. In this context, AI offers significant opportunities (Chen et al., 2022, p. 2) by enabling the rapid creation of realistic artefacts, including written documents, audio narratives, and simulated interactions.

This article presents an experience report on the use of AI tools to generate diverse pedagogical artefacts within a business simulation game (BSG). It illustrates how carefully designed prompts and constrained AI configurations can support immersive learning experiences while preserving the instructor's central role in supervision and decision-making.

The remainder of this paper is organised as follows. The first section outlines the theoretical background. The second section describes the intervention setting. The third section presents the research methodology. The subsequent section reports the results. The paper concludes with a discussion of the findings and final remarks on the role of artificial intelligence in Business Simulation Games.

# Background

*Business Simulation Games and Experiential Learning*

Games can be used as a dedicated medium (*teaching with games*), a repurposed medium (*teaching through games*), or as a subject of study (*teaching about games*) (Pötzsch et al., 2023, p. 350). Business Simulation Games (BSG) are typically dedicated media aiming to simulate managerial situations in order to enhance engagement and make learning more concrete (Ben-Zvi, 2010). Simulations are not a recent phenomenon, as the first business simulations were already used in the United States in the 1950s (Faisal et al., 2022, p. 2), and earlier research indicates that similar approaches were employed as early as the 1930s in the USSR (Gagnon, 1987). Simulations differ from serious games. Serious games are video games used in professional contexts with the objective of enhancing engagement and immersion (Allal-Chérif & Makhlouf, 2016), whereas simulations are more commonly used in academic settings (Faisal et al., 2022). In the Information Systems (IS) field of research, simulations aim to make IS learning more concrete through practical application that complements theoretical lectures and case studies (Ben-Zvi, 2010). For instance, BSG are used in the context of ERP learning, such as SAP, to simulate order processing, delivery management, and the definition of differentiated pricing strategies (Labonte-LeMoyne et al., 2017).

*Artificial Intelligence in Simulation-Based Learning*

The literature documents several contributions of Artificial Intelligence (AI) in educational contexts. For instructors, AI improves the execution of recurrent or repetitive tasks (Labadze et al., 2023). Previous research on the use of AI in education indicates that student benefits mainly concern three areas: learning assistance, personalised learning experiences, and skills development (Labadze et al., 2023).

The use of AI in education is not entirely new; prior studies have already explored its potential and limitations (Chen et al., 2022; Labadze et al., 2023; Larson et al., 2024). However, the intensive use of AI by both students and instructors represents a more recent phenomenon. Importantly, AI should not be viewed as a substitute for instructors; rather, it enhances their capabilities and supports pedagogical activities. AI should be viewed as a cognitive collaborator (Park, 2026).

The main limitations of AI relate to the reliability of certain responses. Unless explicitly configured, AI systems will not state that they do not know an answer. Instead, they may generate what is commonly referred to as *hallucinations* (Larson et al., 2024). Consequently, mastery of a well-defined body of knowledge appears essential in order to avoid such pitfalls (Labadze et al., 2023). However, the notion of hallucination is becoming increasingly less relevant as AI systems have made substantial progress in terms of information reliability (Bengio et al., 2024). In practice, erroneous or fabricated outputs are less a structural property of AI than the result of two factors: the quality of the training data on the one hand, and the response rules imposed on the system on the other. When AI tools are explicitly configured to acknowledge uncertainty and to refrain from responding in the absence of reliable sources the risk of hallucination is largely mitigated. In this sense, AI can be considered a disciplined system rather than an autonomous source of error (Bengio et al., 2024).

The more critical risk may therefore lie elsewhere. If students systematically rely on AI-generated narratives derived from the same underlying datasets, even if these datasets are extremely large, this may limit their exposure to alternative interpretations and perspectives. Analysing heterogeneous events through a single, standardised informational lens may ultimately lead to a form of narrative homogenisation (Park, 2026). Especially regarding Western-centric cultural norms that are embedded in data. Over time, such standardisation risks impoverishing human sensemaking, creativity, and the plurality of interpretations that underpin learning and social development (Larson et al., 2024).

The use of AI in simulation is a well-established phenomenon (Crookall, 2008). Nevertheless, the use of AI in Business Game Simulations (BSG) remains at an early stage. Research combining AI and BSG is still scant (Chen et al., 2022, p. 4), and recent literature review has highlighted this unaddressed research gap (Aliyev et al., 2025, p. 14). Despite this, the potential of AI in this context is considered significant. Prior theoretical contributions indicate that, for learners, the main benefits include the personalisation of learning, real-time interaction, increased engagement and immersion in the learning process, and the modelling of complex or ambiguous situations (Chen et al., 2022). At the same time, previous research points to risks related to student dependence and the lack of critical thinking (Aliyev et al., 2025; Larson et al., 2024).

*Theoretical Framework: the SECI model*

The dynamics of knowledge creation distinguish three types of knowledge: tacit knowledge, explicit knowledge, and phronesis. Explicit knowledge (EK) refers to codified information expressed through a shared and consensual language, making it relatively easy to communicate and transfer, such as scientific publications or financial statements. Tacit knowledge (TK), by contrast, is rooted in action and experience and is therefore difficult to articulate or formalise; typical examples include knowing how to ride a bicycle or recognising a familiar face. Cognitive psychology suggests that procedural memory, which underpins tacit knowledge, is more durable than declarative memory, which underlies explicit knowledge (Kahneman, 2011). A third form of knowledge is phronesis, which is grounded in experience and practical judgement. Phronesis refers to action guided by values, principles, and moral considerations (Nonaka & Takeuchi, 2021, p. 2).

The SECI model identifies four distinct modes of knowledge transmission, based on the interaction between tacit and explicit knowledge (Nonaka & Takeuchi, 2021).

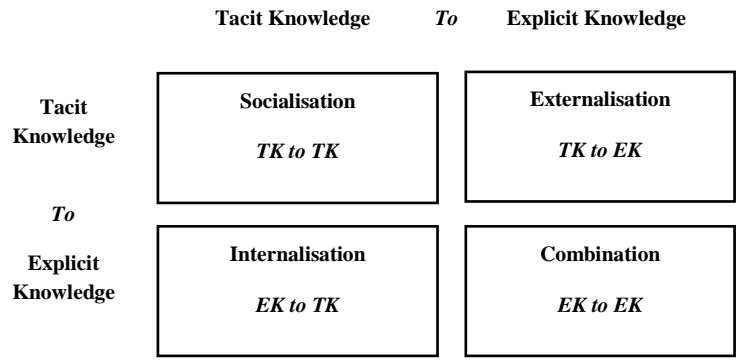

**Figure 1.** The SECI Model adapted from Nonaka & Takeuchi (2021)

*Socialisation* (TK → TK). Transmission based on observation, know-how, and the "unspoken." Often occurs in master-apprentice relationships, where models are invisible and exist as cognitive processes.

*Externalisation* (TK → EK). The process of codifying or translating tacit knowledge into a format that is comprehensible and transmissible. This involves creating a "memory," such as a user manual or a recruitment process map.

*Combination* (EK → EK). Learning based on the mutual exchange of already codified knowledge. Models act as vectors of transmission, such as exchanging scientific articles or conceptual maps, but require mastery of coding rules to be effective.

*Internalisation* (EK → TK). Using codified explicit knowledge to guide practical action. Examples include following a recipe or a maintenance manual; the success depends heavily on the quality of the model and the user's ability to decode it. Table 1 summarises cognitive processes and corresponding model roles.

**Table 1.** Cognitive processes

| Mode | Knowledge Shift | Role of the Model |
|---|---|---|
| Socialisation | Tacit → Tacit | Invisible / Cognitive processes |
| Externalisation | Tacit → Explicit | Memory / Codification |
| Combination | Explicit → Explicit | Transmission vector |
| Internalisation | Explicit → Tacit | Service of action |

# Intervention

*Description of the Business Simulation Game*

The Business Simulation Game (BSG) used in the case report is a commercial product. It aims to turn around a company facing commercial difficulties. It is used by engineering students who have no prior training in management sciences. As a result, most participants are initially unable to read, analyse, and interpret accounting documents or sales and operation planning. The objective of the simulation is to enable students to collaboratively manage a company in a realistic yet controlled environment. I experienced about twenty BSG sessions without AI and two sessions with AI. Each session engages approximately one hundred engineering students.

Engineering students are randomly assigned to teams. Each team ideally consists of seven members, each holding a distinct managerial role: Head of Sales and Marketing, Head of Production, Head of Human Resources, Head of Continuous Improvement and Corporate Social Responsibility, Chief Financial Officer, Management Control Director, and Chief Executive Officer. Roles are distributed by the students themselves. The teaching materials associated with the simulation, on the one hand, and the game documentation and its adaptations, on the other, comprise approximately 200 and 400 typed pages respectively. Mastering the full set of theoretical and practical sources is therefore highly challenging, as students have only one week to familiarise themselves with the documentation. Consequently, students must adopt a clear division of roles and responsibilities in order to successfully complete the assigned mission. This represents a significant challenge: within five days, students are expected to understand how a company operates and to embody a specialised managerial role.

The morning of the first day is dedicated to a question-and-answer session aimed at ensuring a basic understanding of management concepts. Topics include the distinction between commercial, industrial, and service firms; market typologies; reading and interpreting the balance sheet, income statement, cash flow, working capital, and working capital requirements; the principle of double-entry accounting; the notions of cost and expense; cost calculation methods; and an introduction to variance analysis. These foundations are provided to ensure that students are operational during the simulation. At the end of the morning session, students are randomly assigned to a team.

In the afternoon, teams are distributed across different rooms. Each group receives a set of role-specific documents and organises the allocation of functions. Students then review the documentation corresponding to their assigned role. In parallel, workshops on strategy formulation are conducted for sales and marketing directors, based on a strategic management framework encompassing internal analysis, external analysis, and strategy implementation. A workshop on process modelling using Business Process Management and Notation (BPMN) is also organised, enabling students to formalise their organisational processes. Each student is required to design a functional dashboard to support decision-making in their respective role.

From the second day onwards, each student participates in a specific learning by doing workshop dedicated to each business function. Then, each team is required to enter a set of decisions into a commercial simulation system. Each decision set must:

- be grounded in rational analyses based on a functional dashboard;
- remain consistent with the strategy previously defined by the team;
- be taken in accordance with an organisational routine that implements a structured decision-making process.

The simulation system confronts the decisions made by the different teams and positions each company according to multiple performance criteria, including financial, commercial, human-resource, and production-related dimensions.

The simulation is organised into iterative cycles that allow each team to review past results, conduct new analyses, and subsequently make new decisions. This cyclical structure facilitates the establishment of organisational routines. In parallel with this routine, the instructors have defined approximately thirty random events that are introduced to challenge and destabilise the teams. For each event, the AI plays a specific role, which is described in the following sections.

## *Integration of AI Tools in the Simulation*

*AI as an Educative Agent.* At this stage (first day), students are granted access to the AI tool NotebookLM, which contains the full set of game documentation, course materials, and detailed instructions for constructing dashboards. The AI is configured according to the following principles:

- Source completeness verification: the system checks that all sources are activated before responding, in order to prevent circumvention of constraints. If the verification fails, the system displays an error message.
- Source-restricted responses: the AI responds exclusively based on the sources provided in the notebook, in order to prevent hallucinations.
- Language adaptation: the AI responds in the user's language. As the subject matter is new, international students may receive explanations in their native language, avoiding the addition of linguistic difficulty to theoretical complexity.
- Pedagogical redirection: if a question falls outside the scope of the game documentation or course materials, the system responds to contact teaching team.
- Promotion of responsible AI use: if a student requests a full model, a coherent set of decisions, or any output that would effectively replace their work, the system replies that it is not programmed to performed tasks for students. The AI then offers guidance to help students structure their work based on the instructions embedded in the system.

On the first day of the simulation, the student-to-instructor ratio is approximately one instructor for fifty students, making it impossible to respond to all student queries in a timely manner. Consequently, at this initial stage, AI was used as a pedagogical assistant responsible for handling questions of a first level of difficulty. The AI system was also supplemented with frequently asked questions collected from previous editions of the simulation. Rather than replacing the instructor, the AI served as a filtering mechanism, systematically redirecting more complex or context-specific questions to the teaching staff. From the second day the student-to-instructor ratio is approximately one instructor for thirty students.

As shown by research in cognitive psychology, reading primarily relies on System 2 processing (Kahneman, 2011), which requires significant cognitive effort. By using AI to support such rational and analytical operations, students can preserve cognitive resources for the development of procedural knowledge through internalisation processes (Nonaka & Takeuchi, 2021). From the instructor's perspective, explaining concepts also demands substantial cognitive effort. By outsourcing a first level of explanation to AI, instructors can preserve their cognitive resources to address more complex questions and to support students in transforming explicit knowledge into practice through internalisation.

*AI as a Negotiation Agent.* A conversational agent (NotebookLM) was configured to simulate commercial contract negotiations with students during the business simulation. The agent acted on behalf of a subsidiary of a company, and was responsible for selling unbranded drone products, with branding left to the purchasing company.

The negotiation agent operated under a strict rule-based configuration. The AI system was designed to function exclusively when the negotiation rule document was included among the authorised sources, ensuring that all responses remained fully constrained by predefined contractual conditions. Four product categories were available, each associated with standard unit prices and minimum order quantities. Prices were negotiable based on ordered volumes, following predefined quantity thresholds and discount rules.

For each product, the agent applied progressive price reductions when order quantities exceeded specific thresholds. These rules enabled the simulation of realistic volume-based negotiations while preserving transparency and predictability. The AI was not authorised to deviate from these predefined conditions or to invent alternative contractual terms.

During the negotiation process, the agent required students to explicitly commit to both the negotiated unit price and the ordered quantity. Once an agreement was reached, the AI requested the identification code of the student team (e.g., A1) and automatically generated a unique transaction reference number. This reference was then used by the administrative staff to record the transaction.

At the end of each negotiation, the agent produced a structured summary table detailing the order, including product type, quantities, negotiated prices, payment terms, and delivery conditions. Payment was required in full at the time of order, with direct delivery specified as the standard logistics condition. This mechanism enabled students to experience a realistic, rule-constrained commercial negotiation while maintaining full pedagogical control over the process. By outsourcing repetitive tasks to AI such as applying the same rules to each negotiation, which essentially corresponds to a combination process (Nonaka & Takeuchi, 2021) instructors can focus on managing exceptional negotiations that fall outside predefined rules.

*AI as an interactive actor.* Depending on production rates, workplace climate, and the safety measures implemented in each company, the simulator determines the occurrence of occupational accidents. Without real-life enactment, an occupational accident remains merely a numerical value in a table. However, workplace accidents have health, social, and economic consequences. In a business simulation game, it is particularly difficult to simulate an occupational accident realistically. Nevertheless, AI makes it possible to give voice to an employee who has experienced a workplace accident.

To this end, the ElevenLabs platform was used to create a conversational agent tasked with providing a first-person account of the situation within the company. The objective is to demonstrate that occupational accidents have tangible consequences. For instance, labour inspectors may impose penalties or even shut down companies that fail to meet their safety commitments. While the long-term emotional effects of AI on human behaviour need further research, humans still prefer human emotional feedback over AI, even when AI-generated responses are rated as highly empathic (Wenger et al., 2026). Nevertheless, this approach makes the learning experience more vivid and engaging. The agent performed its role very effectively.

*AI in Support of Responsiveness.* In a business simulation, the teaching team must be able to react immediately to the evolving situation of each company. For example, if a company's cash position falls below the overdraft limit authorised by its debt ratio (0.5):

$$\frac{Debts}{Permanent\ capital}$$

The facilitator must promptly notify the company of its financial difficulty. ChatGPT and Gemini are used to formalise personalised messages based on predefined prompts.

In other cases, a company may fall into a growth trap. This situation occurs when the company generates a positive result but is unable to maintain sufficient cash flow due to payment delays from customers. In such cases, the facilitator must propose factoring solutions, allowing the company to exchange accounts receivable for immediate liquidity. Here, AI proves to be a particularly valuable ally, as it can draft highly professional commercial proposals, thereby enhancing the level of immersion in the simulation.

In other situations, AI can be used to enhance the credibility of the discourse and make it more vivid. It is sometimes impossible for the teaching team to achieve such a level of reactivity on its own. For example, within the simulation, certain companies may exploit technologies protected by patents. In such cases, teams found to be in violation of intellectual property rights must be formally notified through a legal notice drafted on behalf of the injured company's lawyer. During the simulation, a company refused the lawyer's proposal for mediation and instead chose to proceed to court. This exercise enables students to understand that, in situations of clear infringement, reaching an agreement is generally preferable to engaging in legal procedure. The objective was to preserve the light-hearted nature of the game while taking seriously the consequences, which have a real impact on the virtual company.

In order to raise students' awareness of the energy challenges currently faced in Europe, where energy consumption restrictions may be imposed during peak demand periods, consumption constraints were introduced through an anxiety-inducing announcement. The message was written and then voiced by an AI agent, recorded as an MP3 file, and broadcast to the students during the simulation.

## Methods

### Research Design and Setting

This study adopts a participant observation within a Business Simulation Game (BSG). Students were informed of the learning, educational, and research objectives. The simulation aims to create a pedagogically credible setting that supports experiential learning in the domain de enterprise management. While, assessment practices in BSG remain underexplored (Vos, 2015), we assessed students formatively using a competency-based evaluation approach. Assessment relied on two complementary dimensions: observable artefacts produced by students during the simulation, and direct observation of student behaviour in authentic decision-making situations (Vos, 2015).

Observable artefacts primarily reflect explicit knowledge (*EK*). The use of AI tools is frequent and aligns with processes of knowledge *Combination*. By contrast, behaviours observed during authentic simulation episodes result from interactions between tacit and explicit knowledge (*EK* to *TK*, *TK* to *EK*, and *TK* to *TK*). In such situations, the role of AI differs, as it involves, *Internalisation*, *Externalisation*, and *Socialisation* rather than merely *Combination* process.

### Research Objective

The research question of this experience report is: *What role does AI play within a Business Simulation Game?* The objective is to explore how AI contributes to the learning process from both student and instructor perspectives. However, this report primarily focuses on the instructor's use of AI to support facilitation, feedback, and pedagogical decision-making during the simulation.

All students participating in the simulation were informed that anonymised data could be used for educational and research purposes. No sensitive personal data were collected during the simulation.

*Simulator and AI Tools*

The simulation is supported by a commercial online platform designed to introduce students to business management through a multidisciplinary perspective. The AI tools used during the simulation included ElevenLabs, Gemini, NotebookLM, and ChatGPT. As this contribution takes the form of an experience report, the study does not rely on a formalised data analysis protocol. Instead, insights are derived from structured observation of learning situations and instructional practices throughout the simulation.

## Results

AI therefore played multiple roles (Table 2) throughout the simulation without ever substituting for instructors. Rather, this research shows that AI can outperform humans in purely combinative knowledge processes. This finding is in line with previous research that hypothesises the ability of AI to provide answers in a short time (Chen et al., 2022, p. 2). Nevertheless, AI has clear limitations that are inherent to its mode of operation. Large Language Models (LLM) rely on textual instructions and generate textual outputs; as such, they are not capable of internalisation, externalisation, or socialisation. These processes require emotional intelligence and procedural cognitive knowledge, which remain fundamentally human. Consequently, instructors are indispensable for enhancing the learning experience. Humans can learn without AI, whereas AI cannot learn without humans. Moreover, AI learning is essentially statistical, while human learning is far more complex: it can be driven by emotions, intuition, embodied experience, and learning through error. AI does not learn through trial-and-error in the experiential and embodied sense characteristic of human learning.

Table 2. summary of AI roles

| Game phase | Role of the AI | Role of the instructor | Contribution of AI | Knowledge type |
|---|---|---|---|---|
| Launch phase | First-level pedagogical assistant | Handling complex or multidimensional questions | Responsiveness/Ubiquity | Combination |
| Launch phase | Multilingual agent | Delivery of pedagogical content | Multilingual support | Combination |
| Planned events | Commercial negotiation agent | Data entry and exceptional negotiations | Responsiveness/Ubiquity | Combination |
| Unplanned events | Interactive agent | Definition of the desired event | Credibility | Combination |
| Unplanned events | Anxiety-inducing voice agent | Definition of the textual corpus to be read | Credibility | Combination |

# Discussion

AI is a powerful tool for summarising, reshaping, or even translating existing explicit knowledge into other forms of explicit knowledge. In other words, AI excels at the combination process (Nonaka & Takeuchi, 2021, p. 2). However, AI is not capable of internalisation, externalisation, or socialisation. *Internalisation* involves transforming explicit knowledge into experience, which requires active engagement and reflection. *Externalisation* refers to converting experience into explicit knowledge a process that inherently requires emotional intelligence and subjective insight. *Socialisation* relies on human interaction, empathy, and shared experience. AI cannot transfer emotions, cannot feel happiness or sadness, and while it can mimic human behaviours or voice patterns (Wenger et al., 2026), this does not generate genuine empathy in human learners.

The role of students and instructors in the learning process is therefore inherently complex. Students must appropriately internalise the explicit knowledge provided by instructors and AI in order to transform it into tacit knowledge. The use of AI alone, without such internalisation, is insufficient for the development of genuine skills. For instance, AI can generate a highly functional dashboard for any business function within a Business Simulation Game. However, if the student does not internalise the underlying logic, the dashboard remains largely unusable and may even create an illusion of understanding without actual mastery. Moreover, dependency on AI remains, at this stage, a hypothesis rather than an empirically established outcome. The long-term effect of technology use over cognition is however a well-known phenomenon (Salomon & Globerson, 1991). In contrast, a student who designs a dashboard independently even if imperfect develops what has been described as *sticky information*, which is more durable and transferable (Von Hippel, 1988).

While the literature on critical thinking is rather fragmented, it is possible to distinguish two forms: the capacity to critically examine and question established knowledge, and the ability to interrogate social norms through a social justice lens (Larson et al., 2024). AI tends to have a natural authority over students, which makes the first form of critical thinking particularly challenging. It is therefore necessary to prevent the uncritical use of AI and the unreflective acceptance of the information it provides. By contrast, with respect to the second form of critical thinking, students appear more naturally critical. Regarding AI's ability to mimic emotions (Wenger et al., 2026), students were able to question the authenticity and social meaning of such representations. When AI-generated narratives were used to simulate a workplace accident, students were not emotionally affected; instead, they often responded with smiles.

The goal is to foster internalisation, as tacit knowledge largely relying on System 1 processing, is more durable and transferable than explicit knowledge, which primarily relies on System 2 (Kahneman, 2011). Instructors play a critical role in internalisation, externalisation, and socialisation processes, which require experience, judgment, and human interaction. By contrast, AI excels in the combination of explicit knowledge. In this perspective, professors and AI should be seen as complementary, each enhancing distinct but interdependent dimensions of the learning process. AI should be viewed as a cognitive collaborator (Park, 2026). This research is consistent with Nonaka and Takeuchi analysis: "*from a knowledge perspective, […] AI does not have tacit knowledge embedded in it; it is all about explicit knowledge*" (Nonaka & Takeuchi, 2021).

Concerning phronesis acquisition, it remains entirely dependent on human instructors. Phronesis refers to practical wisdom grounded in values, principles, and moral judgment (Nonaka & Takeuchi, 2021).

While AI systems can be configured with rules that mimic ethical constraints or normative expectations, they do not possess values, moral reasoning, or responsibility. In the present case, AI tools were explicitly constrained to avoid producing complete solutions for students, for instance by refusing to perform tasks in place of learners. However, such behaviour does not reflect ethical understanding on the part of the AI; it merely results from rule-based textual constraints. By contrast, students are aware that claiming work they did not perform constitutes an ethical violation, highlighting a fundamental asymmetry between rule-following systems and human moral agency. This distinction was further illustrated during the simulation of workplace accidents. For students, such events initially appeared as numerical indicators in spreadsheets. Although AI-generated first-person narratives were introduced to increase realism, some students reacted with detachment or humour, precisely because they knew that the AI had not genuinely experienced suffering. This reaction underscores the limits of artificial empathy (Wenger et al., 2026) and reinforces the idea that phronesis and genuine ethical awareness emerge only through human experience, social interaction, and responsibility (Nonaka & Takeuchi, 2021).

## Limitations and suggestions for further future research

This study has several limitations. First, it is based on an experience report rather than on a rigorous quantitative or qualitative evaluation. As such, the findings should be interpreted as exploratory rather than confirmatory. Future research could address this limitation by systematically measuring student satisfaction, engagement, and patterns of artificial intelligence (AI) use during business simulation activities. As observed in this study, a substantial proportion of students either did not use AI or made only minimal use of it, which warrants closer empirical examination.

Further studies could also investigate the impact of AI use on learners' self-efficacy, drawing on social cognitive theory (Bandura, 2012). In addition, the concept of flow from cognitive psychology (Csikszentmihalyi, 2009) could be mobilized to assess motivation and engagement during simulation-based learning environments (Preuß, 2021). Exploring the interaction between flow states and AI-supported decision-making may represent a promising avenue for future research.

Moreover, future work could examine different configurations and roles of artificial intelligence within simulation games in order to assess their respective effects on learning outcomes and decision-making performance. Finally, systematic observation of instructors during simulation sessions could help identify key pedagogical variables that influence the effectiveness of AI-supported simulations and inform future instructional design.

## Conclusion

This research provides one of the first experience reports combining artificial intelligence with Business Simulation Games. While artificial intelligence is a powerful tool that surpass human capabilities in terms of data processing and rule-based operations (Bengio et al., 2024), this study highlights its clear limitations when compared to human instructors.

Human instructors are able to engage in socialisation, externalisation, combination, and internalisation processes, whereas artificial intelligence is essentially limited to the combination of explicit knowledge (Nonaka & Takeuchi, 2021). Human instructors are essential for tacit knowledge acquisition through experience (Humpherys et al., 2022). Previous research has offered valuable insights into the use of artificial intelligence in Business Simulation Games (Chen et al., 2022). However, empirical evidence remains scarce. Beyond Business Simulation Games, this distinction may help clarify the respective roles of AI and instructors in experiential learning contexts more broadly.

**Acknowledgements**

To be completed

**Funding statements**

This research did not receive any specific grant from funding agencies in the public, commercial, or not-for-profit sectors.

**Competing interests**

The Author(s) declare(s) that there is no conflict of interest.